%% file: main-arxiv.tex
\documentclass{sigchi-arxiv}

\PassOptionsToPackage{pdfpagelabels=false}{hyperref} 

\CopyrightYear{2018}

\toappear{\textcopyright \hspace{1pt} 2018 Copyright held by the owner/author(s)}

\usepackage{balance}       
\usepackage{graphics}      
\usepackage[T1]{fontenc}   
\usepackage{txfonts}
\usepackage{mathptmx}
\usepackage[pdflang={en-US},pdftex]{hyperref}
\usepackage{color}
\usepackage{booktabs}
\usepackage{textcomp}

\usepackage{microtype}        
\usepackage{ccicons}          

\usepackage{subcaption}		  
\usepackage[inline,shortlabels]{enumitem}	

\usepackage{todonotes}

\def\plaintitle{A Study of Material Sonification in Touchscreen Devices}

\def\emptyauthor{}
\def\plainkeywords{Sonification; digital material appearance; multisensory perception.}

\makeatletter
\def\url@leostyle{%
  \@ifundefined{selectfont}{
    \def\UrlFont{\sf}
  }{
    \def\UrlFont{\small\bf\ttfamily}
  }}
\makeatother
\def\pprw{8.5in}
\def\pprh{11in}

\definecolor{linkColor}{RGB}{6,125,233}
\hypersetup{%
  pdftitle={\plaintitle},
  pdfauthor={\emptyauthor},
  pdfkeywords={\plainkeywords},
  pdfdisplaydoctitle=true, 
  bookmarksnumbered,
  pdfstartview={FitH},
  colorlinks,
  citecolor=black,
  filecolor=black,
  linkcolor=black,
  urlcolor=linkColor,
  breaklinks=true,
  hypertexnames=false
}

\begin{document}

\title{\plaintitle}

\numberofauthors{3}
\author{%
  \alignauthor{Rodrigo Mart\'in\\
    \affaddr{University of Bonn}\\
    \email{rodrigo@cs.uni-bonn.de}}\\
  \alignauthor{Michael Weinmann\\
    \affaddr{University of Bonn}\\
    \email{mw@cs.uni-bonn.de}}\\
  \alignauthor{Matthias B. Hullin\\
    \affaddr{University of Bonn}\\
    \email{hullin@cs.uni-bonn.de}}\\
}

\maketitle

\input{macros}

\begin{abstract}
Even in the digital age, designers largely rely on physical material samples to illustrate their products, as existing visual representations fail to sufficiently reproduce the look and feel of real world materials. Here, we investigate the use of interactive material sonification as an additional sensory modality for communicating well-established material qualities like softness, pleasantness or value. We developed a custom application for touchscreen devices that receives tactile input and translate it into material rubbing sound using granular synthesis. We used this system to perform a psychophysical study, in which the ability of the user to rate subjective material qualities is evaluated, with the actual material samples serving as reference stimulus. Our experimental results indicate that the considered audio cues do not significantly contribute to the perception of material qualities but are able to increase the level of immersion when interacting with digital samples.
\end{abstract}

\category{H.5.2.}{Information Interfaces and Presentation (e.g. HCI)}{User Interfaces} {}{}


\input{intro}

\input{related}

\input{experiments}

\input{results}

\input{discussion}

%
%
%
%
%
\balance{}




\bibliographystyle{SIGCHI-Reference-Format}
\bibliography{biblio}

\end{document}

%% file: macros.tex
\definecolor{marine}{rgb}{0.0,0.5,0.5}
\definecolor{orange}{rgb}{1.0,0.5,0.0}
\definecolor{purple}{rgb}{0.2,0.0,0.8}

\definecolor{mygreen}{rgb}{0.2,0.8,0.2}
\definecolor{myred}{rgb}{0.8,0.1,0.1}

\def\ie{i.e.\ }
\def\eg{e.g.\ }

\newcommand\tab[1][1cm]{\hspace*{#1}}
\newcommand{\vect}[1]{\bm{#1}}
\newcommand{\mat}[1]{\mathbf{#1}}

%% file: intro.tex
\section{Introduction}

The shopping experience in our everyday life is determined by various types of interaction with commodities. A consumer's decision-making process in a store, for instance, is based on a multitude of inner rating processes that do not only involve the perception of physical properties through different senses such as sight, hearing or touch, but also an emotional or affective experience. In the context of online shopping and materials in particular, the sensory and emotional bandwidth of interaction with the respective commodity is greatly reduced and mostly limited to a passive visual representation and, occasionally, a textual description. Previous investigations have shown that the lack of a multimodal experience in general, and tactile input in particular, significantly affects the user's capability of assessing physical material properties (\eg softness, flexibility) and developing affective emotions (\eg pleasantness, value) evoked by the product \cite{Citrin2003}. As a logical consequence is desirable to enhance the digital material with additional cues on top of the purely visual user experience. In this regard, auditory cues and sonification techniques have demonstrated to influence significantly the perception of a product quality/efficiency \cite{Spence2006}, compensate the absence of tactile interaction with digital material samples \cite{Martin2015} and increase the feeling of immersion w.r.t. the unimodal visual experience \cite{Ho2013}. Instead of simply triggering a prerecorded audio sample of the interplay with the material, directly allowing customers to interact with the digital material, where audio information corresponding to this interaction is automatically synthesized in real time represents an interesting challenge for the digital commerce.

The goal of this work is the analysis of the effects in the perception of physical and affective material qualities when visual material representations (photographs) are augmented with interactive audio feedback generated as the response to a single finger rubbing motion. Thereby, we employed a granular synthesis approach to build a sonification system that allows to enrich the user interaction with digital material samples through touchscreen devices. We then conducted a user study in which participants rated a set of relevant material qualities across a purely visual condition, two audiovisual conditions (including static pre-recorded sound and the interactive sonification system) and a full-modal condition, in which they were able to interact with the actual specimen. The experimental results were examined by means of the degree of correlation between participants, the analysis of the perceptual space spanned by the material qualities, the performance of each condition in a material classification task and the examination of the elapsed time per experimental condition.

The key findings from this set of analyses are the following:
\begin{itemize}
\item The addition of rubbing material sounds as such does not significantly improve the perception of material qualities, although the overall material experience is not compromised by the presence of auditory cues.
\item The presence of interactive audio increases the level of immersion of the users when interacting with digital materials.
\end{itemize}

In light of these findings, we provide plausible explanations for the experimental results and suggest profitable directions of future research.


%% file: related.tex
\section{Related work}

This section provides a summary of relevant investigations in the areas of multimodal perception of materials as well as sonification and sound synthesis of material interactions.

\subsection{Multimodal perception of materials.}
The majority of the research in product perception has been focused on the visual modality. Nevertheless, auditory cues have demonstrated to have a significant influence in the perception of product quality/efficiency (including electric toothbrushes, cars or foodstuff), and to be able to provide semantic signatures to a certain brand (\eg the breaking sound of a `Magnum' or the opening of a `Schweppes' bottle) \cite{Spence2006}. Like any other product, the perception of materials is inherently multimodal and, with it, several strands of research have been conducted to investigate how the interplay between different senses shapes the perception of textures, materials and objects. An extensive review of the perception of textures regarding touch, vision and hearing is provided by Klatzky and Lederman \cite{Klatzky2010}, in which texture is understood as a perceptual property that characterizes the structural details of a surface.

There are not many approaches that focus on the investigation of purely acoustic material perception. Klatzky et al.~\cite{Klatzky2000} analyzed the relationship between material perception and variables that govern the synthesis of impact sounds. Their results indicate the importance of a shape-invariant decay parameter in the perception of the material of which an object is made, while the frequency content plays also an important part. In a related fashion, Giordano and McAdams \cite{Giordano2006} studied the human performance when identifying materials from impact sounds. Interestingly, they concluded that listeners performed well with respect to the gross categories, but their performance degraded for materials belonging to the same category.

Beyond the human performance in classifying materials, also the ability to infer concrete material qualities has received particular attention. Fleming et al.~\cite{Fleming2013} conducted a set of experiments to investigate the interactions between material classification and quality judgments. A high degree of consistency between these two assignments was detected, indicating that they facilitate one another by accessing the same perceptual information. The multisensory nature of the communication of material qualities has been further explored by Mart\'in et al.~\cite{Martin2015}, where the authors employed contact and stroking material sounds to complement the visual stimuli. Their results demonstrate the strong linkage between the auditory channel and the haptic perception to a point in which sound is capable of biasing the visual judgment of concrete qualities. In addition, Fujisaki et al. \cite{Fujisaki2015} examined how a set of physical and affective qualities of wood are evaluated in three different modalities of vision, audition and touch, and observed that all three senses yield somewhat similar representations. Lastly, The qualities related to aesthetic perception of materials also play an important role in the decision-making process. In fact, strong connections have been observed, for instance, between the assessed smoothness of tactile textures and their perceived pleasantness \cite{Etzi2014}.


\begin{figure*}
\centering
	\begin{subfigure}{0.15\textwidth}
	\includegraphics[width=1\linewidth]{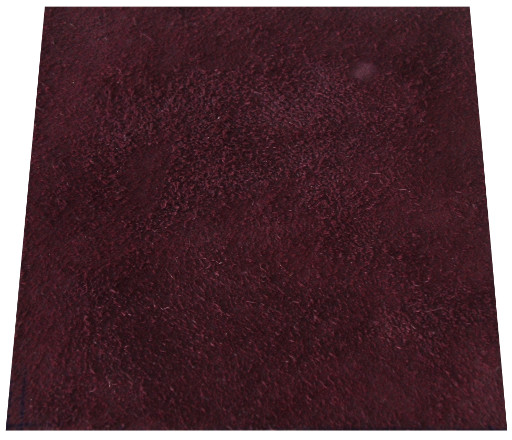}
	\end{subfigure}
	\begin{subfigure}{0.15\textwidth}
	\includegraphics[width=1\linewidth]{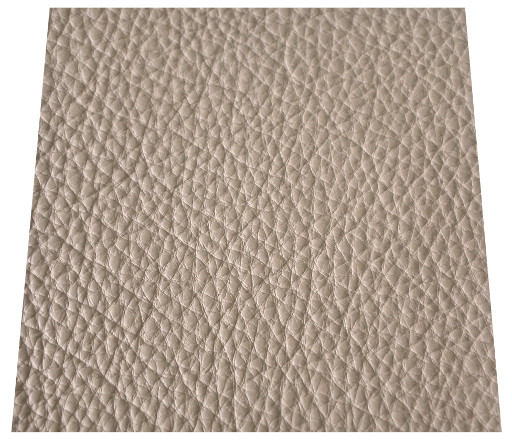}
	\end{subfigure}
	\begin{subfigure}{0.15\textwidth}
	\includegraphics[width=1\linewidth]{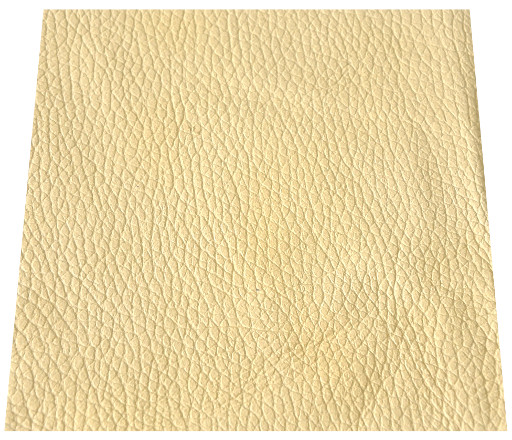}
	\end{subfigure}
	\begin{subfigure}{0.15\textwidth}
	\includegraphics[width=1\linewidth]{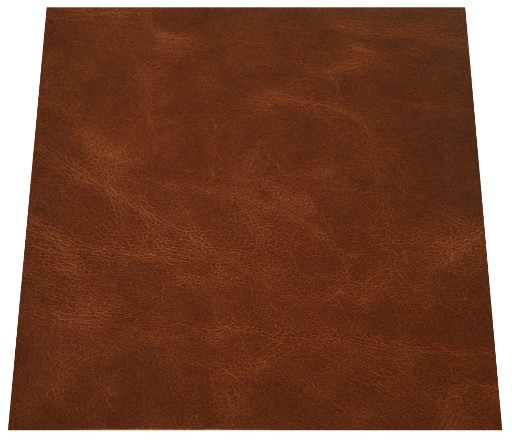}
	\end{subfigure}
	\begin{subfigure}{0.15\textwidth}
	\includegraphics[width=1\linewidth]{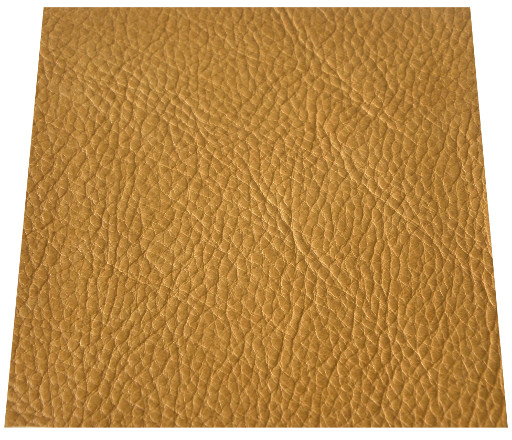}
	\end{subfigure}
	\\
	\begin{subfigure}{0.15\textwidth}
	\includegraphics[width=1\linewidth]{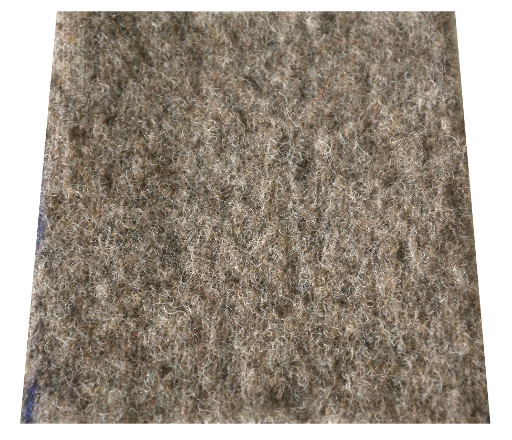}
	\end{subfigure}
	\begin{subfigure}{0.15\textwidth}
	\includegraphics[width=1\linewidth]{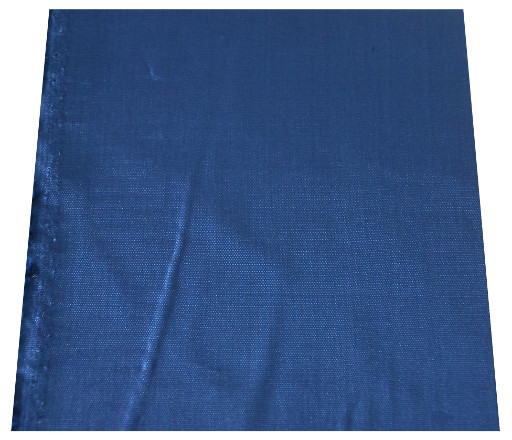}
	\end{subfigure}
	\begin{subfigure}{0.15\textwidth}	
	\includegraphics[width=1\linewidth]{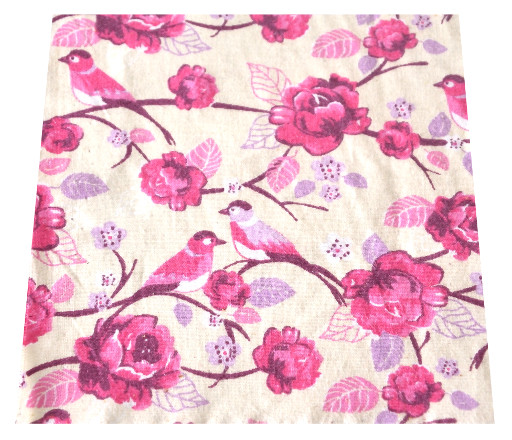}
	\end{subfigure}	
	\begin{subfigure}{0.15\textwidth}
	\includegraphics[width=1\linewidth]{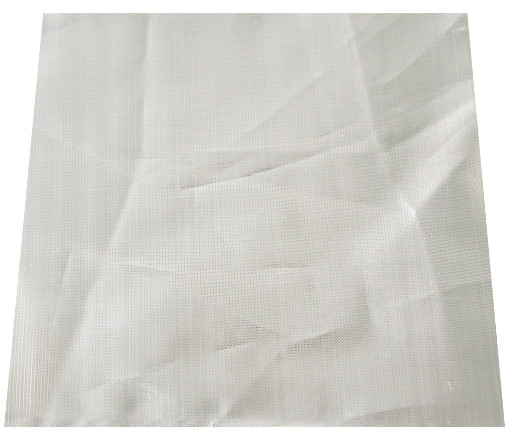}
	\end{subfigure}	
	\begin{subfigure}{0.15\textwidth}
	\includegraphics[width=1\linewidth]{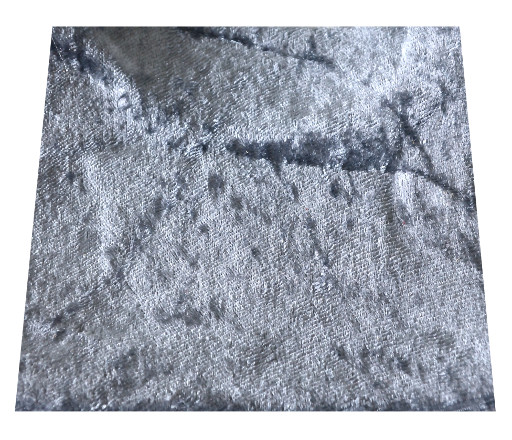}
	\end{subfigure}
\caption{Pictures of the leathers materials (upper row L1--L5) and fabrics (lower row F1--F5) as displayed in our study.}
\label{figure:materials}
\end{figure*}

\subsection{Sonification and sound synthesis of material interactions.}
The synthesis of a multitude of sounds, from artificial to natural and pure musical ones, has important applications in movie sound effects, video games, virtual reality, multimedia or in art installations. An exhaustive survey on the predominant digital audio analysis and synthesis methods has been presented by Misra and Cook~\cite{Misra2009}. The former includes a taxonomy which introduces the most suitable synthesis approaches for each sort of sound and addresses the value and adaptability of the family of granular synthesis methods in the generation of audio textures. In fact, although granular methods have been mainly used in the creation of soundscapes, their possible range of applications include the synthesis of acoustic instruments, pitched sounds, speech, singing voice, and contact sounds from virtual surfaces when bouncing, being broken or scraped \cite{Barrass2002}. Belonging to the same family of techniques as granular synthesis, concatenative synthesis has been utilized in the context of simulating the particular sounds that certain materials produce. An et al.~\cite{An2012} developed a motion-driven algorithm that is able to synthesize cloth sounds for a wide range of animation scenarios. Their technique avoids expensive physics-based synthesis but still produces plausible results. However, it requires a certain amount of manual intervention and does not achieve interactivity.

In the context of visualization, information about the scene is represented in terms of shapes of varying sizes with attached color information and used to create pictures we can look at. Sonification is the equivalent concept translated to the sense of hearing, that is, the synthesis of non-speech audio to convey certain information. In this regard, the use of synthesized material sound has a large range of applications in sonification systems, which may be used to overcome limitations in representing tactile properties of digital objects and materials, among other purposes. An early investigation from Guest et al.~\cite{Guest2002} evaluated how tactile textures are perceived under real-time manipulation of touch related sounds. Their study indicates that the frequency content of textural sounds represents the dominant factor for such sort of interactions as e.g. attenuating high frequencies caused the textures to be perceived smoother. Later, Tajadura-Jim\'{e}nez et al.~\cite{Tajadura-Jimenez2014} explored the ability of a sound-based interaction technique to alter the perceived material of which a touched surface is made. Their granular sonification algorithm reproduces samples (grains) with three different frequency levels where the grain selection is guided by the finger pressure on a wooden surface. With their results, the authors determined that increasing sound frequency alters either the surface perception (colder material) and the emotional response (increased pressure and touch speed). Finally, by using textural sounds in the context of a retail clothing application, Ho et al.~\cite{Ho2013} demonstrated that the simple addition of realistic auditory feedback to the unimodal visual experience favors the feeling of immersion, which becomes evident in longer interaction times with the product and also willingness to pay a higher price for it.

The present investigation establishes, to our knowledge, a novel and interactive approach to material sonification with consumer hardware and the first assessment of its effects on the perception of physical and affective material qualities. We hence arranged a user experiment to evaluate such effects in comparison to additional stimuli, including the actual material samples. The description of our experimental setup is introduced in the following section.

%% file: experiments.tex
\section{Experimental design}

The key elements of our experimental design are given by the considered visual stimuli, the auditory stimuli, the actual material samples and the description of the user study. In the following, we provide details regarding each of these components.

\subsection{Visual stimuli}

In the scope of this research, we explore the perception of physical and affective material qualities for two semantic classes (leathers and fabrics), which are commonly available in retailing websites. For this purpose, we have chosen ten material samples, each of them with an approximate size of $ 120 \times 120 \text{~mm}^2 $, with nearly flat geometry to avoid possible sources of visual variability and sound artifacts. Restricting our selection to these two concrete, well-known classes allows us to keep the study and its conclusions manageable. We then situated each specimen on a bright background under natural illumination and took a picture using a digital camera (Nikon1 J5, resolution $ 5568 \times 3712 \text{~pixels} $) located at approximately $ 200 \text{~mm} $ from the sample under a slight angle. The resulting images were corrected regarding white-balance and scaled to match the resolution of the final device (see Section \ref{subsection:procedure}). The characteristic borders of each specimen were additionally cropped, since they have been demonstrated to provide supplementary information to the material texture that could bias the visual stimuli \cite{Martin2017}. The resulting photographs are displayed in Figure \ref{figure:materials}.

\subsection{Auditory stimuli: acquisition of material sound.}

In order to record the touch-related sounds arising from the interaction with the selected materials (\ie brushing them with the fingertip), we assembled a setup composed by a piece of polyurethane foam with the size of $400 \times 400\,\text{~mm}^2 $, on which the sample was placed. The actual recording step was carried out in an acoustically isolated chamber using a stereo pair of small-membrane condenser microphones in X/Y setting located about $200\,\text{~mm}$ away from the sample and facing towards it. The contact sounds were generated by gently rubbing the material's surface creating random trajectories while increasing the velocity of the movement for roughly a minute. Other than trimming, no further post-processing was applied to the audio. Such recordings were employed as the static audio in one of our experimental conditions (see Section \ref{subsection:procedure}). 

\begin{figure*}
\centering
	\begin{subfigure}{0.33\textwidth}
	\includegraphics[width=1\linewidth]{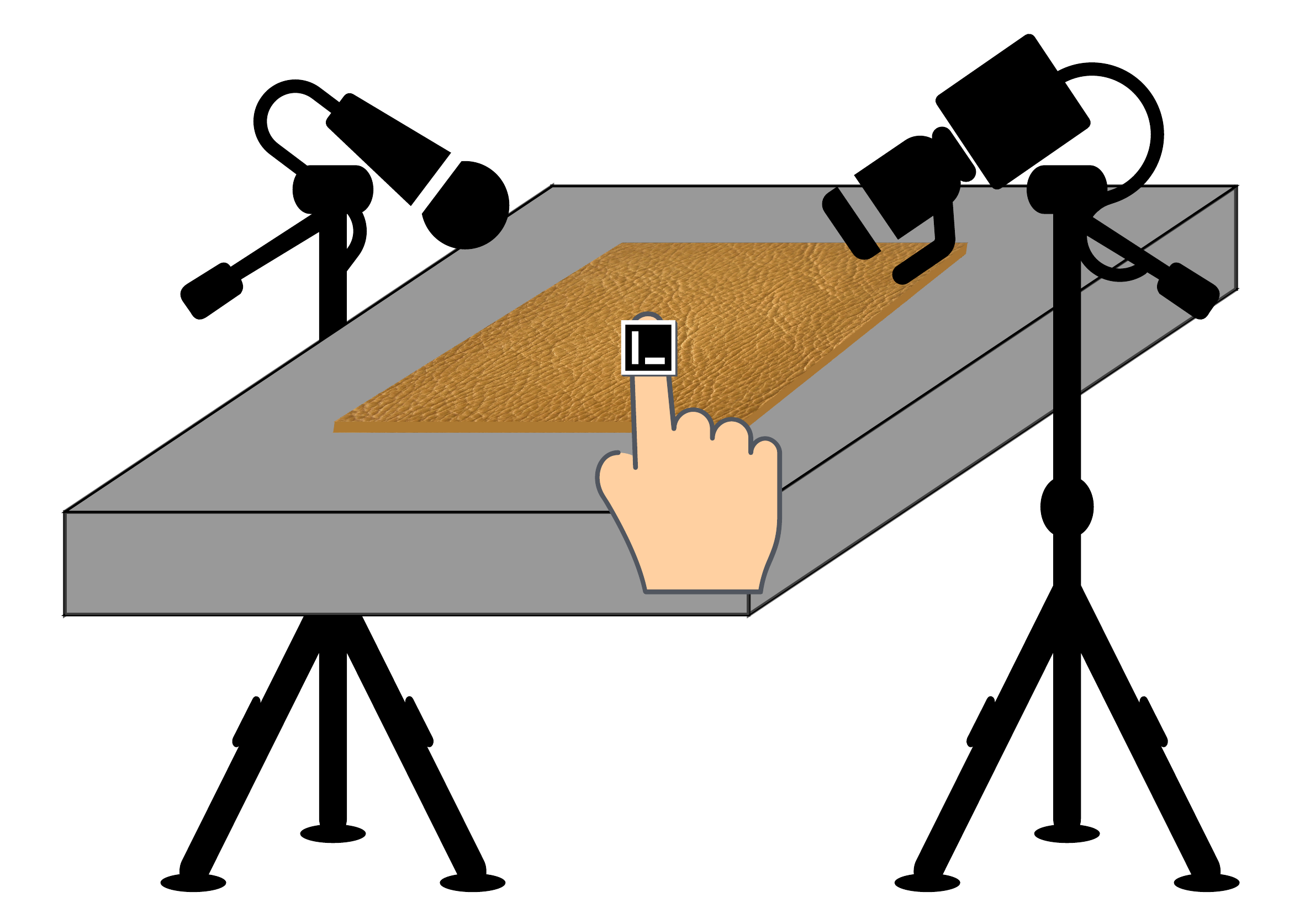}
	\caption{Acquisition setup.}
	\label{subfigure:recording-setup}
	\end{subfigure}
	\begin{subfigure}{0.33\textwidth}
	\includegraphics[width=1\linewidth]{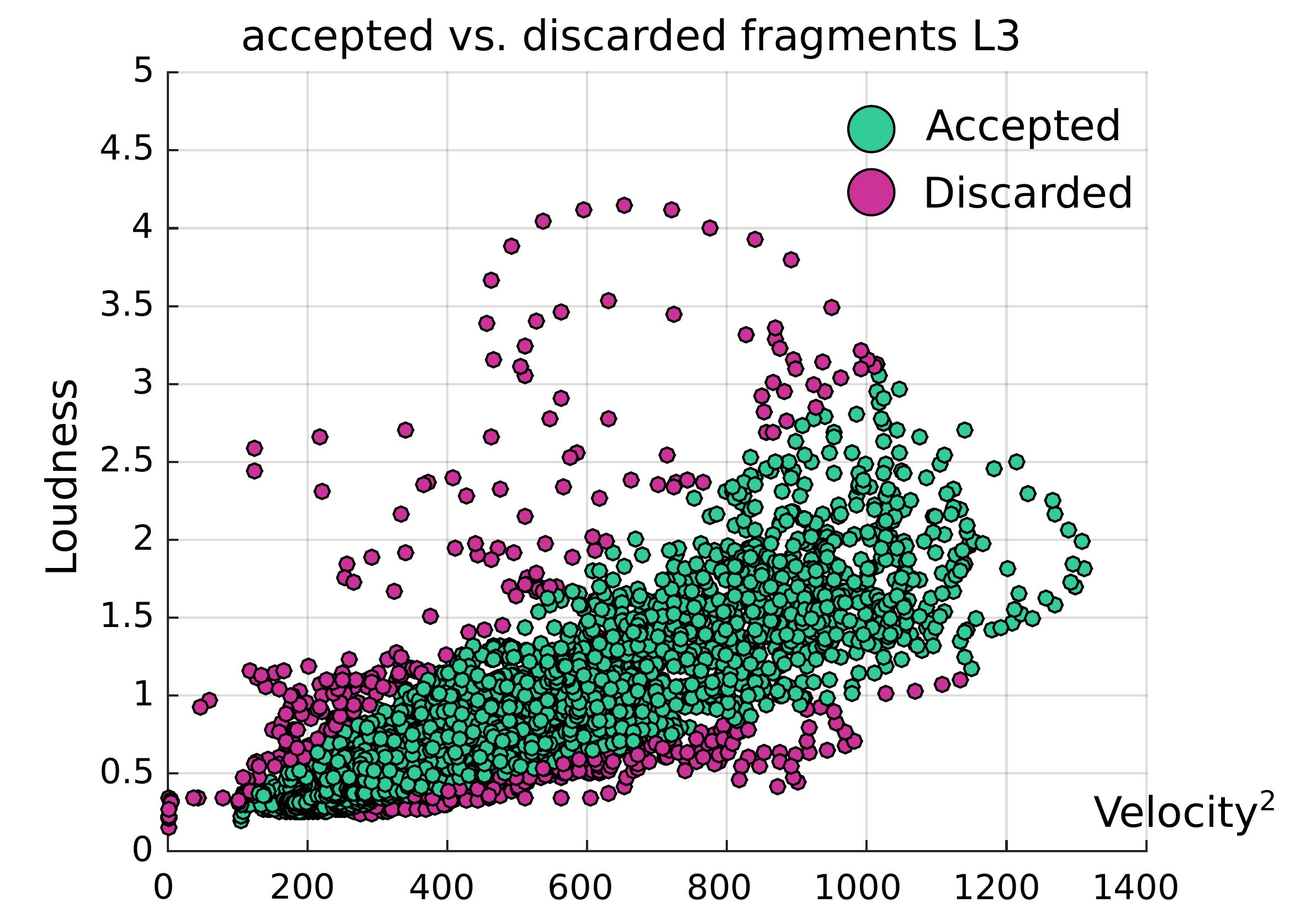}
	\caption{Fragment loudness vs. squared velocity.}
	\label{subfigure:discarded}
	\end{subfigure}
	\begin{subfigure}{0.33\textwidth}
	\includegraphics[width=1\linewidth]{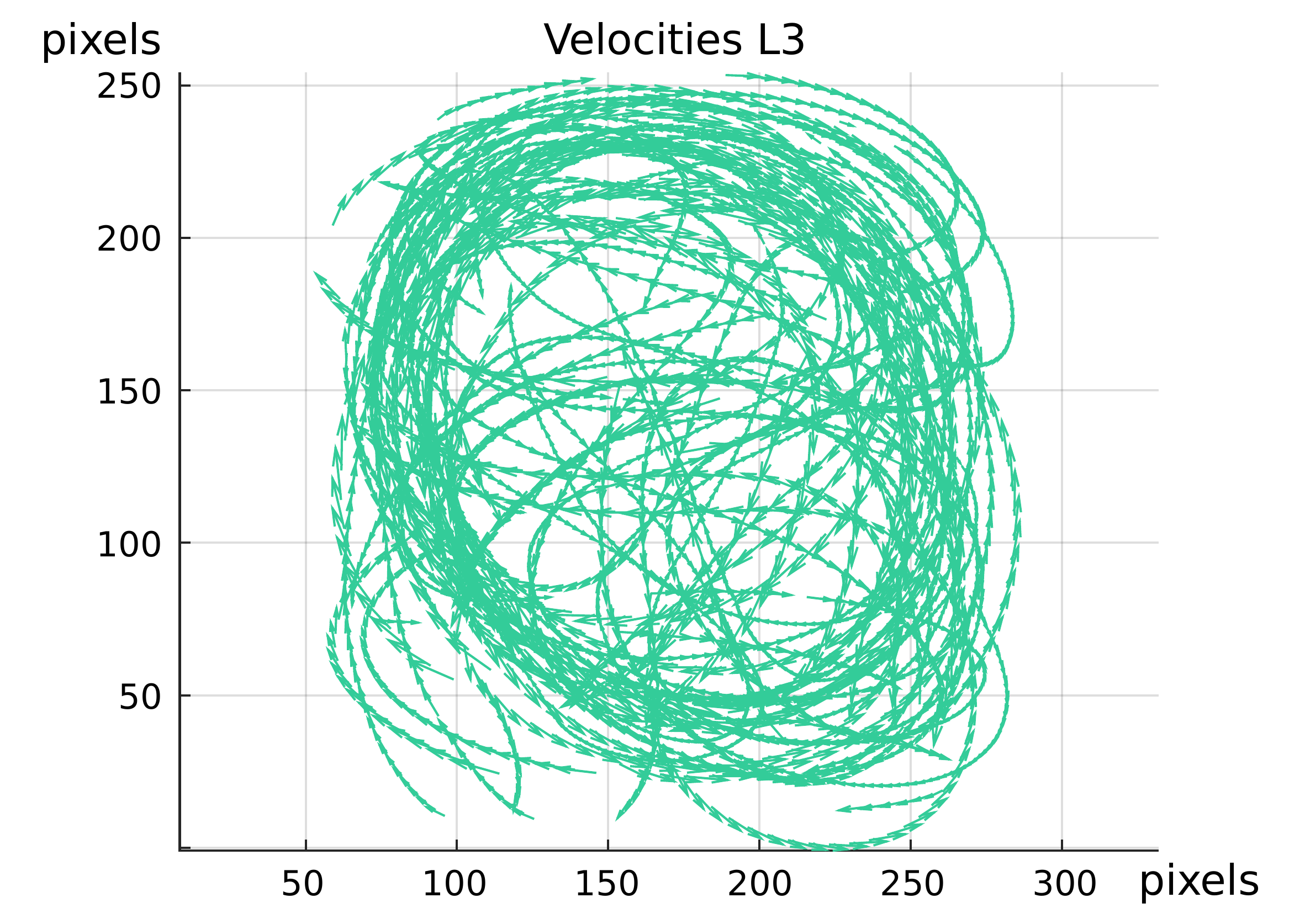}
	\caption{Velocities of accepted fragments.}
	\label{subfigure:velocities}	
	\end{subfigure}	
	\\
	\vspace*{6pt}
	\begin{subfigure}{0.28\textwidth}
	\includegraphics[width=1\linewidth]{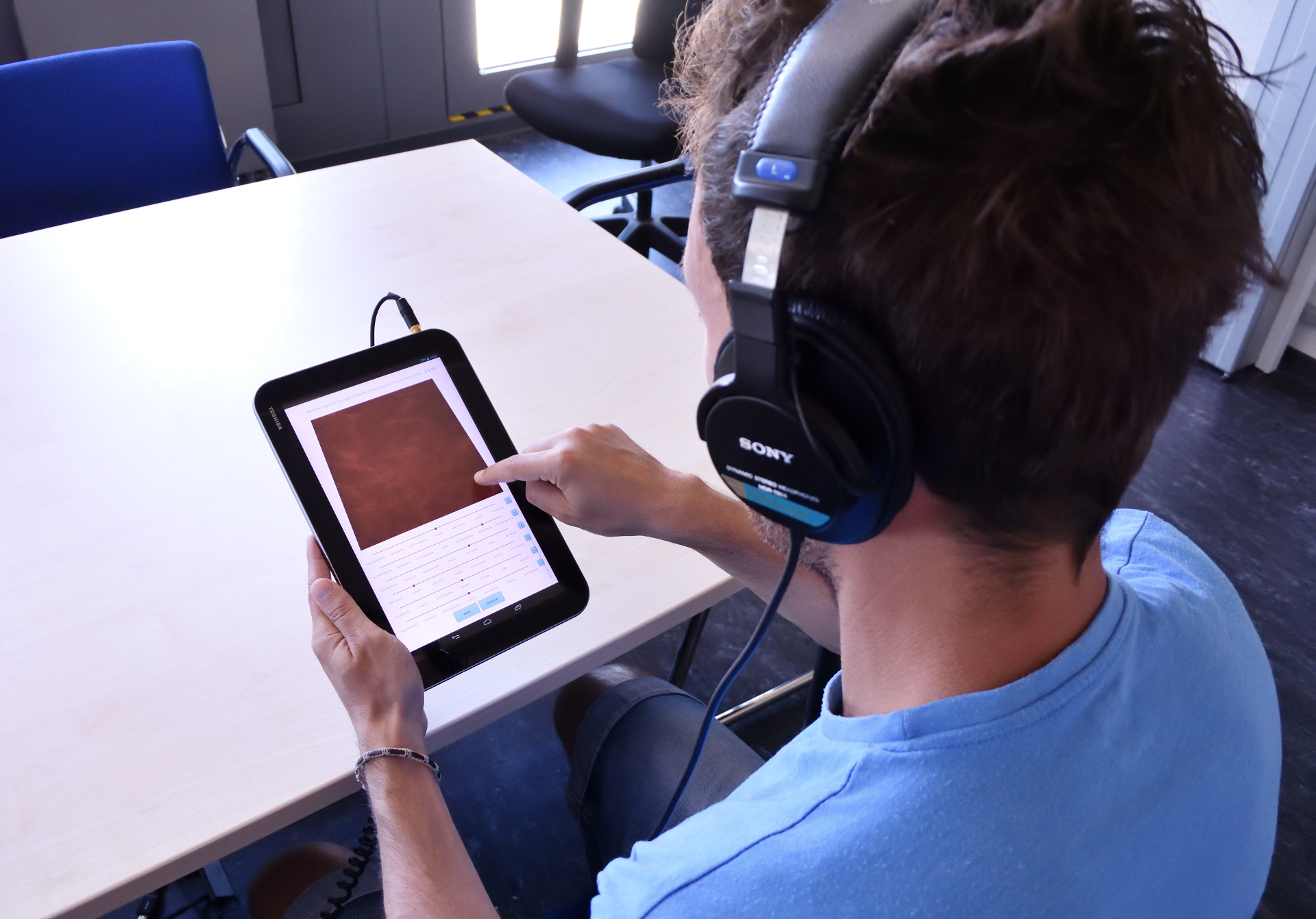}
	\caption{Participant during the study.}
	\label{subfigure:interaction}
	\end{subfigure}
	\hspace{20pt}
	\begin{subfigure}{0.42\textwidth}
	\includegraphics[width=1\linewidth]{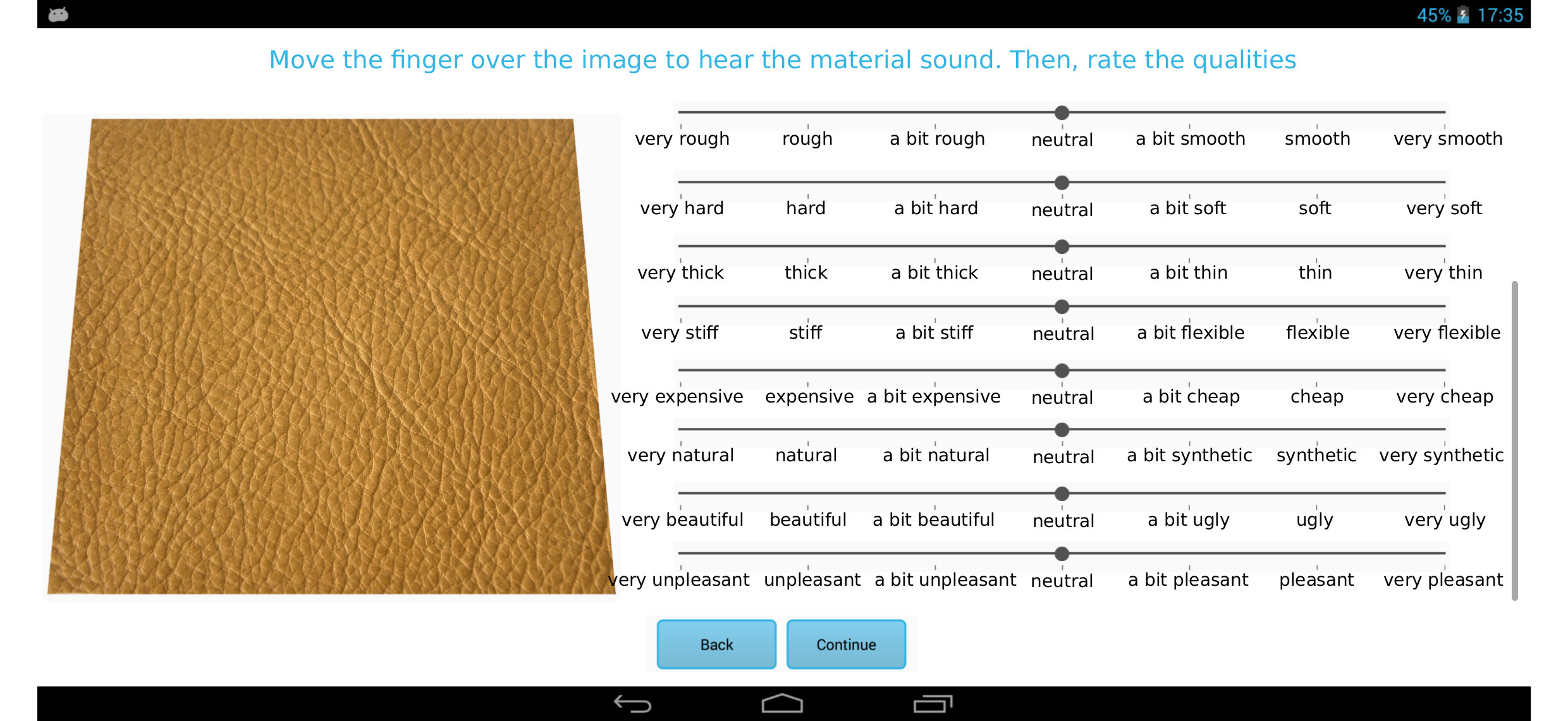}
	\caption{Elements of the user interface.}	
	\label{subfigure:screenshot-example}
	\end{subfigure}
\caption{Illustration of the relevant steps from the synthesis process and the experimental study, including the acquisition of touch-related material sound (a), pre-processing phase for a concrete material L5 (b) and (c), a picture of a participant during the study (d) and the elements composing the user interface (e).}
\end{figure*}

To later guide our sound synthesis method, we annotated the resulting signal with the position of the finger during the interaction. We achieved this by attaching a fiducial marker \cite{GarridoJurado2014} to the nail of the interacting finger, which we tracked using a machine vision camera (Point Grey GS3-U3-23S6M-C Grasshopper). An illustration of the complete setup is depicted in Figure \ref{subfigure:recording-setup}. After basic analysis of the video data (marker tracking, trajectory smoothing, numerical differentiation), we thus obtained 2D finger velocity data $ \vec v_{i}$ at 100 samples per second (see Figure \ref{subfigure:velocities}) along with the 48\,kHz stereo audio clip $s(t)$.


\captionsetup[subfigure]{font=footnotesize}

\subsection{Auditory stimuli: synthesis of material sound.}

To deliver material contact sounds in response to the user's input in real-time, we developed a sonification algorithm based on granular synthesis. This family of techniques has been broadly used in many applications due to their flexibility \cite{Misra2009}, including the synthesis of material contact sounds \cite{Barrass2002}. The moderate computational load, in comparison to physically-based approaches, represents a decisive aspect as it allows to perform the synthesis on consumer hardware (such as tablet computers) at interactive rates, thus facilitating the present study. To this end, we employed the open-source audio processing language \texttt{Pure Data} (Pd)\footnote{\texttt{Pure Data} (Pd) is an open source visual programming language for multimedia. For more information and resources we refer to the corresponding webpage: \url{http://libpd.cc}}, which is the most widespread audio synthesis environment under Android \cite{Brinkmann2012}, as an embeddable library (\texttt{Libpd}).

We divided the recorded sound clip $s(t)$ into \emph{fragments} with a length of 480 samples, corresponding to the spacing of velocity samples (10\,ms). Each such fragment was annotated by velocity values $ \vec v_j $ (see above) and its root-mean-square (RMS) loudness $ a_j $, noting that the loudness of a fragment roughly scales with the square of the corresponding velocity. To avoid artifacts when re-synthesizing fragments into a new audio stream, we discarded those fragments that were unusually soft or unusually loud for the given velocity. We identified such outliers by computing the ratio $ \alpha_j = a_j/|v_j|^2 $ for each fragment, and then removed those fragments whose ratio was below the $ 5^{th} $ percentile or beyond the $ 95^{th} $ percentile (see Figure \ref{subfigure:discarded}). The remaining set of annotated fragments constitute the input to the sonification system.

During user interaction with the tactile device, the touch interface measures the user's finger velocity $\vec v_\textrm{in}$ on the screen. The granular synthesis uses this value to retrieve suitable sound fragments according to a distance metric that considers the velocity and loudness of the $ j^{th} $ fragment:
\begin{equation}
d_{j} = \sqrt{\left\Vert 
\vec v_\textrm{in}-\vec v_j \right\Vert_2^2 + (|v|^2_\textrm{in}-a_j/\hat\alpha)},
\end{equation}
where $\hat\alpha$ is the mean of the ratio $ \alpha_j $ across all fragments. To ensure variation, we follow a standard practice in granular synthesis by retrieving not only the single closest hit for the given query velocity $ \vec v_\textrm{in} $, but the $ k=25 $ closest fragments instead. With the goal of real-time operation in mind, this $ k $-nearest-neighbor search is implemented using a balanced binary space partitioning (BSP) tree \cite{Fuchs1980}. The synthesis algorithm randomly selects one of these fragments and ``freezes'' it for the upcoming few iterations to avoid repetition artifacts. The fragment is then extended into a longer \emph{grain} by incorporating its $n=28$ neighbor fragments in the input sound clip. Finally, the grain is concatenated and blended (cross-faded) with the previous grain into a continuous audio output. 

Overall, this rather simple system is capable of producing a smooth and interactive stream of contact sound that is free of disturbing artifacts (transitions, repetition) that are otherwise typical for granular synthesis. A major drawback of the mobile platform remains the somewhat long system latency of approximately 500\,ms, inherent to the utilization of the \texttt{Libpd} library under Android.

\subsection{Real materials}

During the progress of the experiments, participants were also asked to evaluate the actual materials samples (full-modal interaction). Instead of using the same specimens utilized during the audiovisual stimuli acquisition, smaller portions of the same samples (approximately $ 70 \times 70 \text{~mm}^2 $) were handed to the users. With this, we avoided damaging the originals during the interactions and facilitated the scalability of the experiment.

\subsection{Task and procedure} \label{subsection:procedure}

\begin{table}
\footnotesize
\centering
\begin{tabular}{l|p{2.5cm}|l}
	\toprule
	\textbf{Tactile} & \textbf{Visual} & \textbf{Affective} \\ 
	\midrule
	rough--smooth & shiny--matte & expensive--cheap \\
	hard--soft & bright--dark & natural--synthetic \\
	thick--thin & transparent--opaque & beautiful--ugly \\
	stiff--flexible & homogeneous--\linebreak heterogeneous & unpleasant--pleasant\\
	\bottomrule
\end{tabular}
\caption{Opposite-meaning quality pairs, grouped by category.}
\label{table:qualities}
\end{table}

Inspired by previous investigations \cite{Fleming2013,Martin2015,Martin2017} we gathered a collection of 24 adjectives describing material appearance. At the same time, these adjectives were organized into 12 opponent pairs which were assigned to either the tactile, visual or affective category, depending on the nature of the physical or emotional interaction that best reveals them (see Table \ref{table:qualities}). In order to rate this set of qualities across our multimodal stimuli, we made use of single stimulus ratings in which the participants assessed each quality pair under study on a 7-point Likert scale, represented with a slider with values ranging from\,$ -3 $ to $ 3 $. The values along the scale were consistently labeled with a term indicating the intensity of the stimuli (e.g., very hard, hard, a bit hard, neutral, a bit soft, soft and very soft). The user study was conducted using tablet computers (Toshiba Excite Pro 10.1, resolution $ 2560 \times 1600 \text{~pixels} $) and a set of headphones (Sony MDR-7506) running a custom Android application which connects with the Pd module. The complete experimental setup can be seen in Figure~\ref{subfigure:interaction} and the user interface is depicted in Figure~\ref{subfigure:screenshot-example} with greater detail. The procedure itself consisted of four presentations or conditions, in which material stimuli were presented in random order to the participants along with the 12 slider widgets. The four conditions that compose the study are the following:

\begin{itemize}
\item \textit{Visual condition (VI)}, where the stimuli are photos taken from real materials.
\item \textit{Static Audiovisual condition (SA)}, where the photos were complemented with prerecorded audio from the material.
\item \textit{Dynamic Audiovisual condition (DA)}, where the photos were complemented with interactive sound generated by our sonification system. This means that real-time contact sound is played back upon tactile interaction with the images on the device.
\item \textit{Full-modal condition (FM)}, consisting of physical material samples that were given to the participants so that they could interact with them.
\end{itemize}

Since the interaction with the real samples could bias the realization of the visual and audiovisual conditions, the full-modal presentation was constrained to be the fourth and final one, while the order of the remaining conditions was randomized. 19 participants took part in the experiment (12 females, mean age $ 27.08 $; 7 males, mean age $ 28.57 $). All the participants were na\"ive to the goals of the experiment, provided informed consent, reported normal or corrected-to normal visual and hearing acuity and were compensated economically for their cooperation. From this experiment, a total of $ 19 \times 10 \times 12 \times 4 = 9120 $ responses were collected and evaluated.

%% file: results.tex
\section{Results}

In order to investigate the effects of our sonification system on the perception of material qualities, we evaluate the correlation between the participants' ratings, the dimensionality of the spanned perceptual space per experimental condition, the performance in a classification task based on the material quality ratings and the time elapsed by the participants in each experimental condition.

\subsection{Inter-participant correlation}

\begin{figure*}
\centering
	\begin{subfigure}{0.24\textwidth}
	\includegraphics[width=1\linewidth]{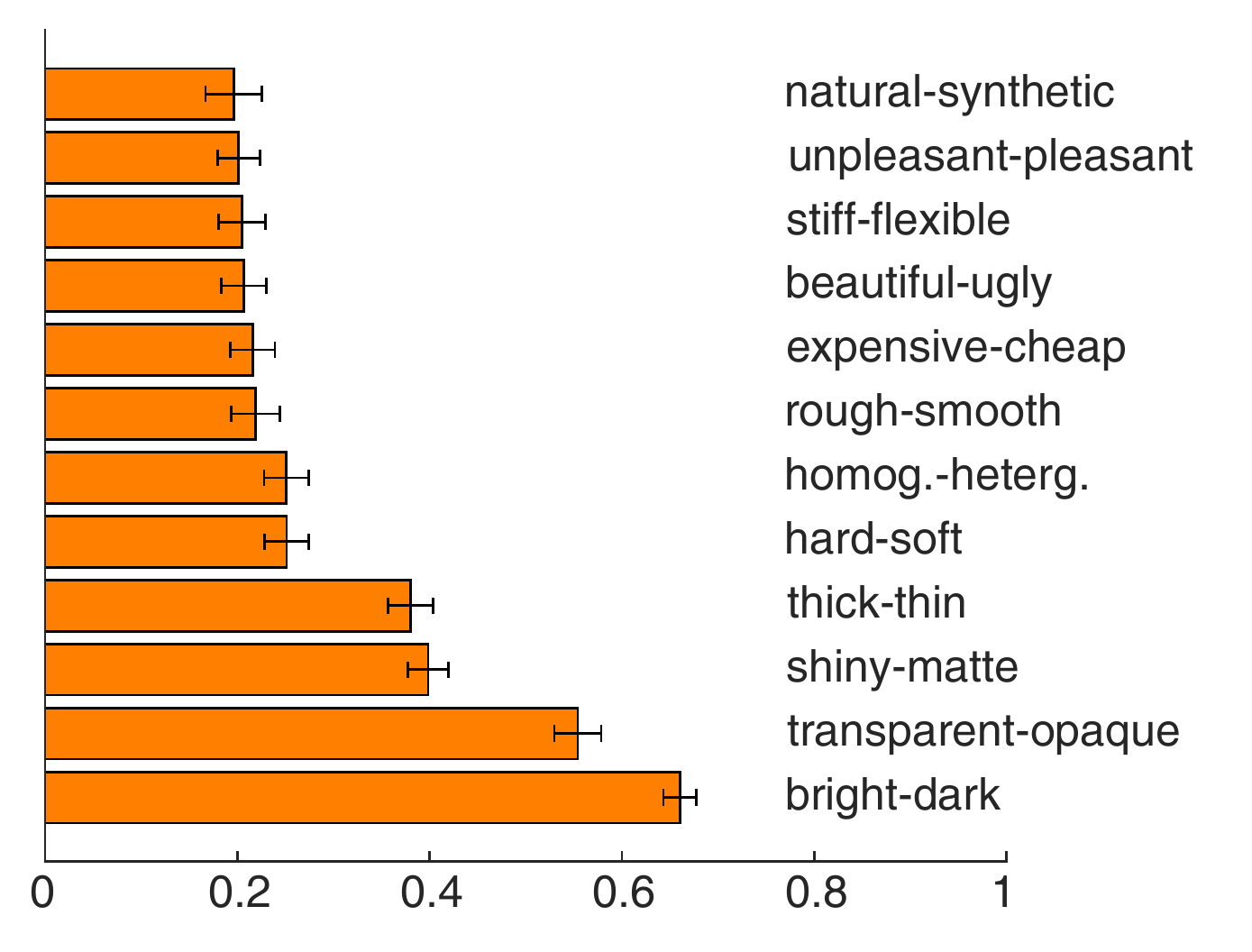}
	\caption{Visual}
	\end{subfigure}
	\begin{subfigure}{0.24\textwidth}
	\includegraphics[width=1\linewidth]{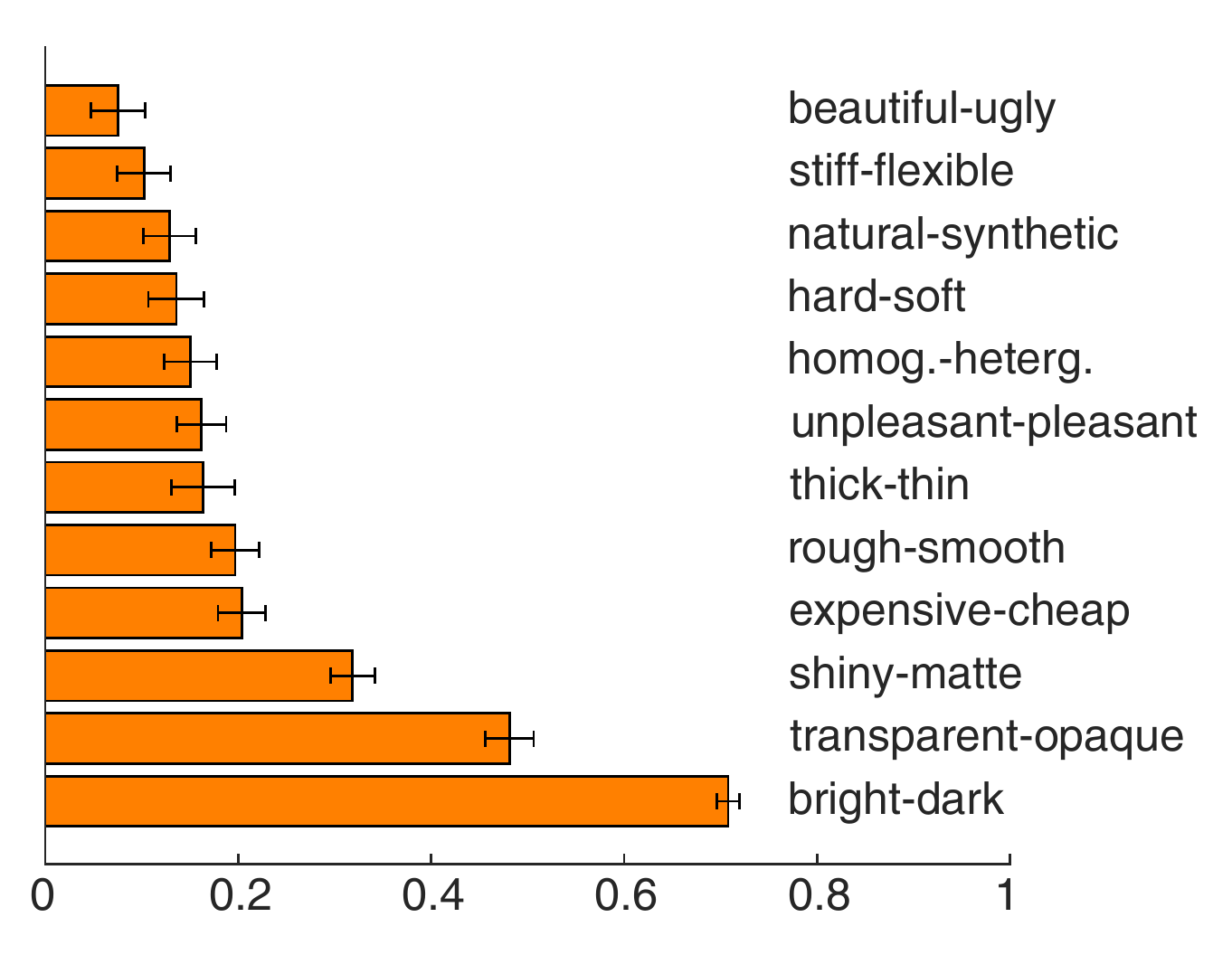}
	\caption{Av. Static}
	\end{subfigure}
	\begin{subfigure}{0.24\textwidth}
	\includegraphics[width=1\linewidth]{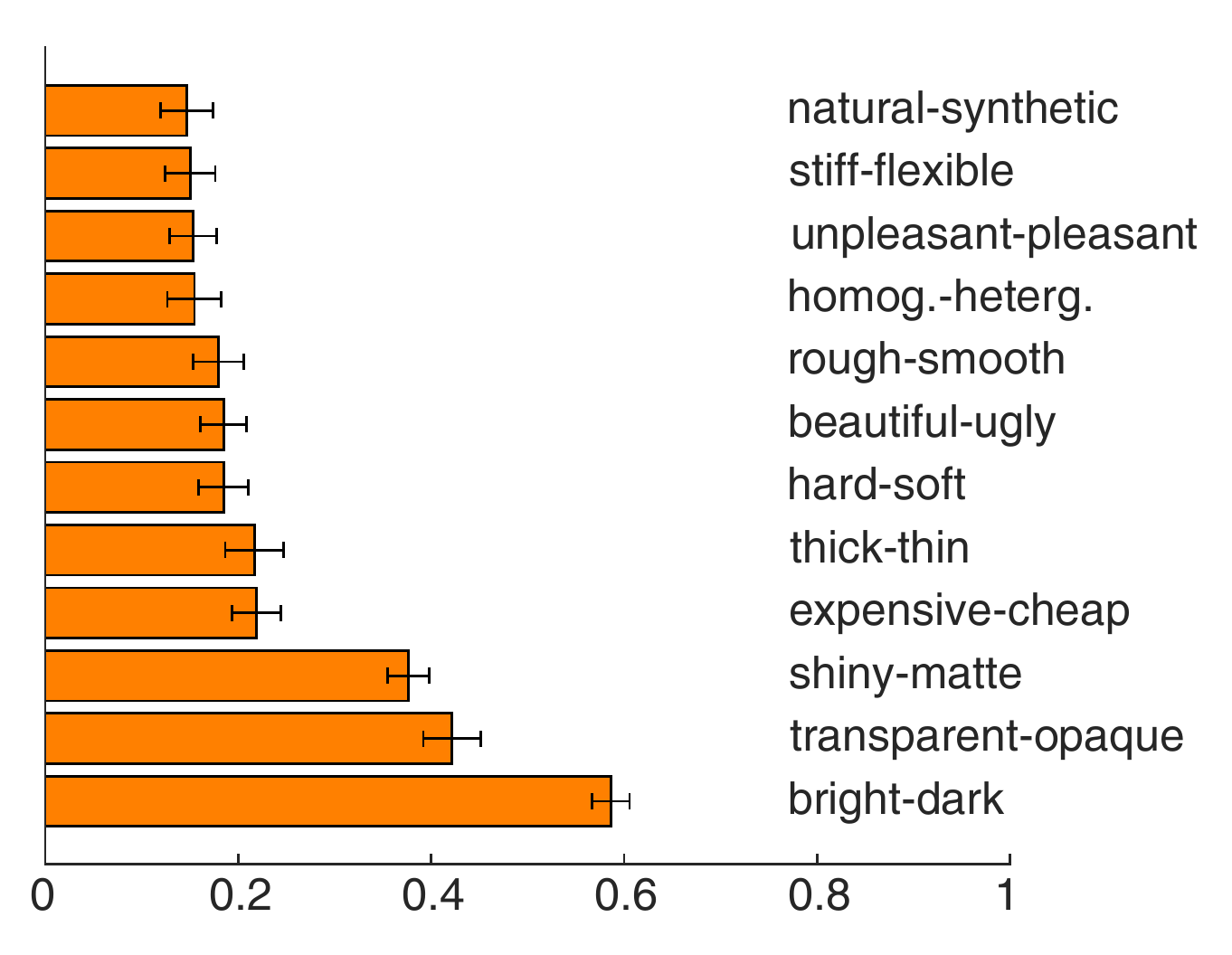}
	\caption{Av. Dynamic}
	\end{subfigure}	
	\begin{subfigure}{0.24\textwidth}
	\includegraphics[width=1\linewidth]{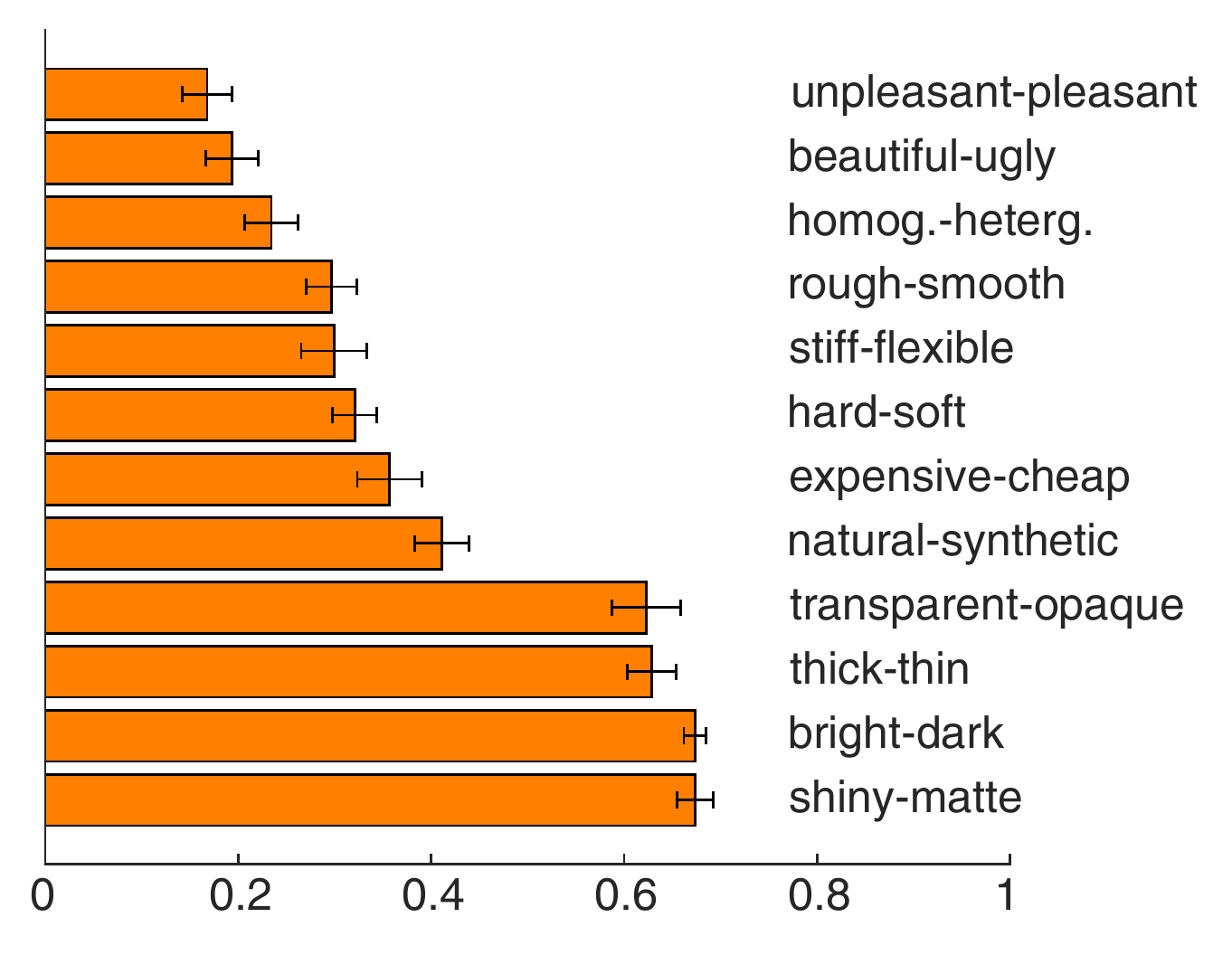}
	\caption{Full-modal}
	\end{subfigure}
\caption{Average inter-participant correlation per property, grouped by condition and sorted in ascending order w.r.t.\ the correlation. Note that the differences between the visual and the two audiovisual conditions are relatively small and how the full-modal condition presents significantly higher correlation values.} 
\label{figure:subject-correlation-prop}
\end{figure*}

Due to the diverse collection of materials and qualities considered in this investigation, we first provide an analysis of the level of agreement between the participants' ratings for the given stimuli. To that end, we computed the inter-participant correlation coefficients for each condition and quality pair, over all materials. The hypothesis assumption is that the higher the correlation coefficient, the better a specific quality would be represented by the condition at hand. Contrarily, if such quality is not well depicted, the users would have to infer it using their imagination, resulting in a lower degree of agreement. Figure~\ref{figure:subject-correlation-prop} illustrates the resulting correlations in ascending order, separated by experimental condition.

\captionsetup[subfigure]{font=footnotesize}

The largest coefficients presented by the visual condition are those corresponding to the pairs ``bright--dark'', ``transparent--opaque'', ``shiny--matte'' and ``thick--thin'', which are properties mostly categorized as visual. Likewise, both audiovisual condition exhibit the largest correlation values for the pairs ``bright--dark'', ``transparent--opaque'' and ``shiny--matte''. However, the ``thick--thin'' dimension shows a much lower value for the DA and particularly the SA condition in comparison to the VI presentation. Allegedly, the proposed rubbing/stroking sounds employed are less suitable for communicating this particular dimension and seem to mislead users' judgments of the material thickness. This is further implied by the correlation values for the full-modal condition, where this pair shows again a significant level of agreement. Another interesting observation is that the user agreement for the tactile qualities as well as the pair ``beautiful--ugly'' is slightly higher in the DA condition when compared to the static audio (SA). Albeit being a promising trend, the effect is not significant enough to draw categorical conclusions. In general, the correlation values ($ R $) and ordering are quite similar for the three digital conditions VI ($ \hat{R}_{VI}=0.32 $), SA ($ \hat{R}_{SA}=0.27 $) and DA ($ \hat{R}_{DA}=0.28 $), and follow a comparable ordering as the full-modal condition ($ \hat{R}_{FM}=0.46 $).

Although in principle this analysis is analogous to the inter-participant correlation from Mart\'in et at. \cite{Martin2015}, the results are not directly comparable, as neither the stimuli employed nor the quality set are entirely identical in both experiments. Specifically, the pictures from the visual presentation in the former investigation display the distinctive borders of materials, which are known to be a powerful discriminator \cite{Martin2017}. Furthermore, the authors included tapping impact sounds in their audiovisual condition, which possibly allowed the inference of additional material information. Taking this into consideration, the larger discrepancy between both studies concerns to the resulting correlation coefficients for the ``hard--soft'' dimension, where the present experiment exhibits considerably lower values for the SA, DA and FM conditions. We conclude that tapping sounds provided decisive cues to assess the hardness of the material. Moreover, the presence of relatively hard paper materials in \cite{Martin2015} probably established an upper bound for this concrete quality, which is not present when solely considering leathers and fabrics.

\subsection{Dimensionality of the perceptual space}

In the previous section, we examined the four different conditions through the correlation between participants, observing little effects between the conditions VI, SA and DA. To further explore this insight, we analyze the dimensionality of the perceptual space spanned by the perceptual qualities. For this purpose, we averaged the ratings over all participants and performed principal component analysis (PCA) for each experimental condition on the mean data. The resulting factor loadings of the first three principal components as well as the explained and accumulated variance are shown in Table \ref{table:factor-loadings}, separated by condition.

A detailed inspection of the coefficients exposes that the first principal components (PC1) in all three digital conditions (VI, SA, DA) are dominated by the visual qualities (``shiny--matte'', ``bright--dark'' and ``transparent--opaque''), which account for most of the variation in the users' ratings. Additionally, VI exhibits somewhat large values in the tactile dimensions, especially for the ``thick--thin'' pair, which are not so evident in SA or DA. This is in accordance with the correlation values reported earlier. Furthermore, the second PC of all three conditions is commonly determined by the ``rough--smooth'' quality and the affective properties, while PC3 has diverse values for each condition. In contrast, the first PC of the full-modal condition is driven by a mixture of qualities (``thick--thin'', ``shiny--matte'', ``transparent--opaque'' and ``natural--synthetic''), while the second PC explains much less variance and is dominated by the roughness, shininess, heterogeneity and the affective qualities.

\input{table-pca}

\begin{figure*}[!ht]
  \centering
  \includegraphics[width=1.0\linewidth]{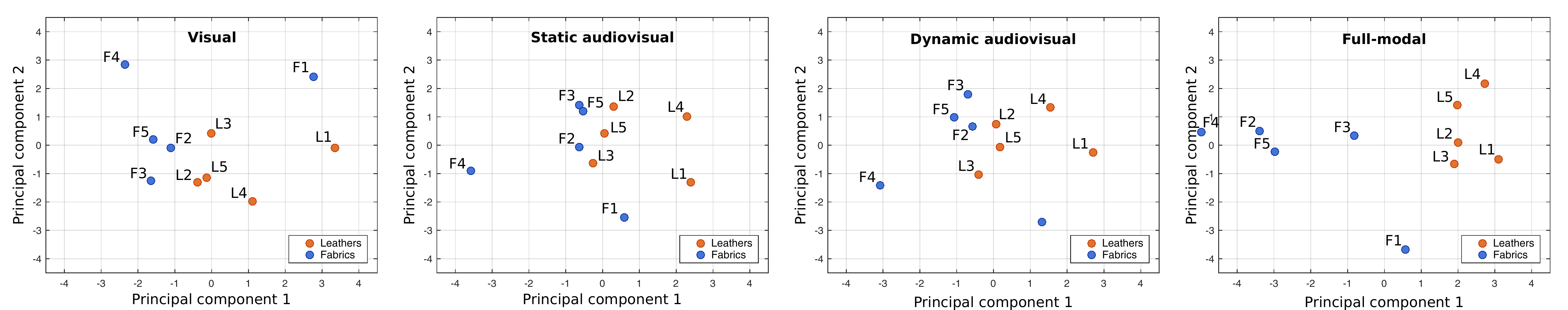}
  \caption{Distribution of the samples in the first two PCs. Circles represent the projected positions of individual material samples (10 in total) in the subdimensional space for each condition. The sample distribution presented by VI, SA and DA are rather identical, while the variance in FM corresponds mostly to the first PC.}
  \label{figure:pca-2dim}
\end{figure*}

When considering the cumulative variance, two dimensions are able to explain $ 69.78\% $, $ 67.29\% $, $ 67.01\% $ and $ 78.51\% $ of the variance for the VI, SA, DA and FM conditions respectively. Therefore, projecting the factor loadings into a $ 2 $-dimensional space seems to be a plausible and easy-to-visualize option to analyze the distribution of the user data (see Figure \ref{figure:pca-2dim}). By inspecting the arrangement of materials in the subdimensional space, we observe that the sample distributions presented by VI, SA and DA are quite similar (PC2 in VI is upside-down). Meanwhile, the variance in FM is primarily accumulated in the first PC, which allows a smooth clustering of the two material classes.

Lastly, we applied procrustes analysis \cite{Gower1975} to compare our experimental conditions $ C_{s} \in \{VI_{s}, AS_{s}, AD_{s}\} $ against the full-modal space ($ FM_{s} $) resulting from PCA, which is taken as a reference. We took the 12-dimensional space spanned by the qualities into account and used the minimized sum of square errors (SSE) to measure the goodness of the mapping. The fitting values for all three $ C_{s} $ are, again, extremely similar with a relatively low error, where $ VI_{s} $ achieves the best result ($ SSE=0.214 $) closely followed by $ SA_{s} $ ($ SSE=0.240 $) and $ DA_{s} $ ($ SSE=0.242 $). From this analysis we conclude that the considered visual (VI) and audiovisual (SA, DA) stimuli are capable of effectively transmitting information about our set of materials and qualities. However, the addition of these specific audio cues, no matter whether in terms of their static form or the sonification system, does not contribute with significant additional information to simple photographs.

\subsection{Material classification}

The previous analysis facilitated the understanding of the ability of the considered stimuli to depict a set of relevant material qualities through the agreement level between subjects and the subdimensional space that they span. Previous studies have demonstrated that humans access the same perceptual information about materials while performing both material categorization and quality rating tasks \cite{Fleming2013}. Keeping this in mind and considering that our stimuli consist of two classes of materials, this section attempts to clarify to what extent such classes can be by predicted based on the participants' ratings. Concretely, we aim at answering the following questions:

\begin{enumerate}[1)]
	\item Which is the classification performance of the experimental conditions (VI, SA, DA) in comparison to the FM condition?
  	\item Do any of the utilized sound cues facilitate the discrimination between leathers and fabrics?
  	\item Which set of considered qualities allows a better material classification?
\end{enumerate}

For this purpose, we trained a binary Support Vector Machine (SVM) classifier to obtain a model which employs the user ratings of the twelve perceptual features to predict the material class to which each sample belongs. We then conducted leave-one-out cross validation tests per user and material sample. Additionally, we performed the same analysis using the ratings from each perceptual category individually as predictors (tactile, visual and affective) as well as all three combinations of them. The accuracy across each condition and group of qualities is provided in Table \ref{table:svmc-accuracy}.

Regarding the first inquiry, performing the classification task on the ratings from the FM condition results into considerably higher accuracies (at least above $ 72\% $) in comparison to the rest of the conditions. This outcome is to be expected, as judging the real material samples will always allow a more confident quality assessment as images or sounds. With respect to question number two, the classification results of the SA and DA conditions exhibit lower values as the visual condition for all the set of predictors considered, by a slight margin. In light of this results, we assume that the addition of rubbing sounds does not help in distinguishing leathers and fabrics, and that the proposed sonification system has additional value over static sounds only when tactile-related predictors are included. Concerning the third question, using all twelve perceptual qualities as predictors yields by far higher accuracies. The table shows also how the visual predictors, alone or in combination with other features, have more discriminating power than tactile or affective qualities. More interestingly and less anticipated is the fact that the use of affective predictors lead to higher accuracies than tactile ones, when the conditions VI, SA and DA are considered. However, tactile features provide better discrimination in the full-modal case, since the participants were able to actually touch the specimens.

\begin{table*}
\centering
\renewcommand{\arraystretch}{1.2}
\begin{tabular}{l|c|c|c|c|c|c|c}
	\cmidrule[\heavyrulewidth]{2-8}
	 & \multicolumn{7}{c}{\textbf{Set of Predictors}} \\
	\midrule
	\textbf{[\%]} & \textbf{All} & \textbf{(T)actile} & \textbf{(V)isual} & \textbf{(A)ffective} & \textbf{T+V} & \textbf{T+A} & \textbf{V+A} \\
	\midrule
	\textbf{VI} & $75.8\%$ & $63.2\%$ & $67.9\%$ & $65.3\%$ & $70.5\%$ & $71.6\%$ & $73.7\%$ \\
	\textbf{SA} & $72\%$ & $56.1\%$ & $66.1\%$ & $65.1\%$ & $68.8\%$ & $61.4\%$ & $69.8\%$ \\
	\textbf{DA} & $66.3\%$ & $60.5\%$ & $61.6\%$ & $62.6\%$ & $66.3\%$ & $69.0\%$ & $69.5\%$ \\
	\textbf{FM} & $89.5\%$ & $78.4\%$ & $82.6\%$ & $72.6\%$ & $89.5\%$ & $84.2\%$ & $89.5\%$ \\
	\bottomrule
\end{tabular}
\caption{Accuracy [\%] of a SVM material classifier based on the perceptual qualities. Each row represents the accuracy for the considered experimental condition, while the columns describe the set of qualities used as predictors.}
\label{table:svmc-accuracy}
\end{table*}

\subsection{Level of immersion}

\begin{figure}
  	\centering
  	\includegraphics[width=0.6\linewidth]{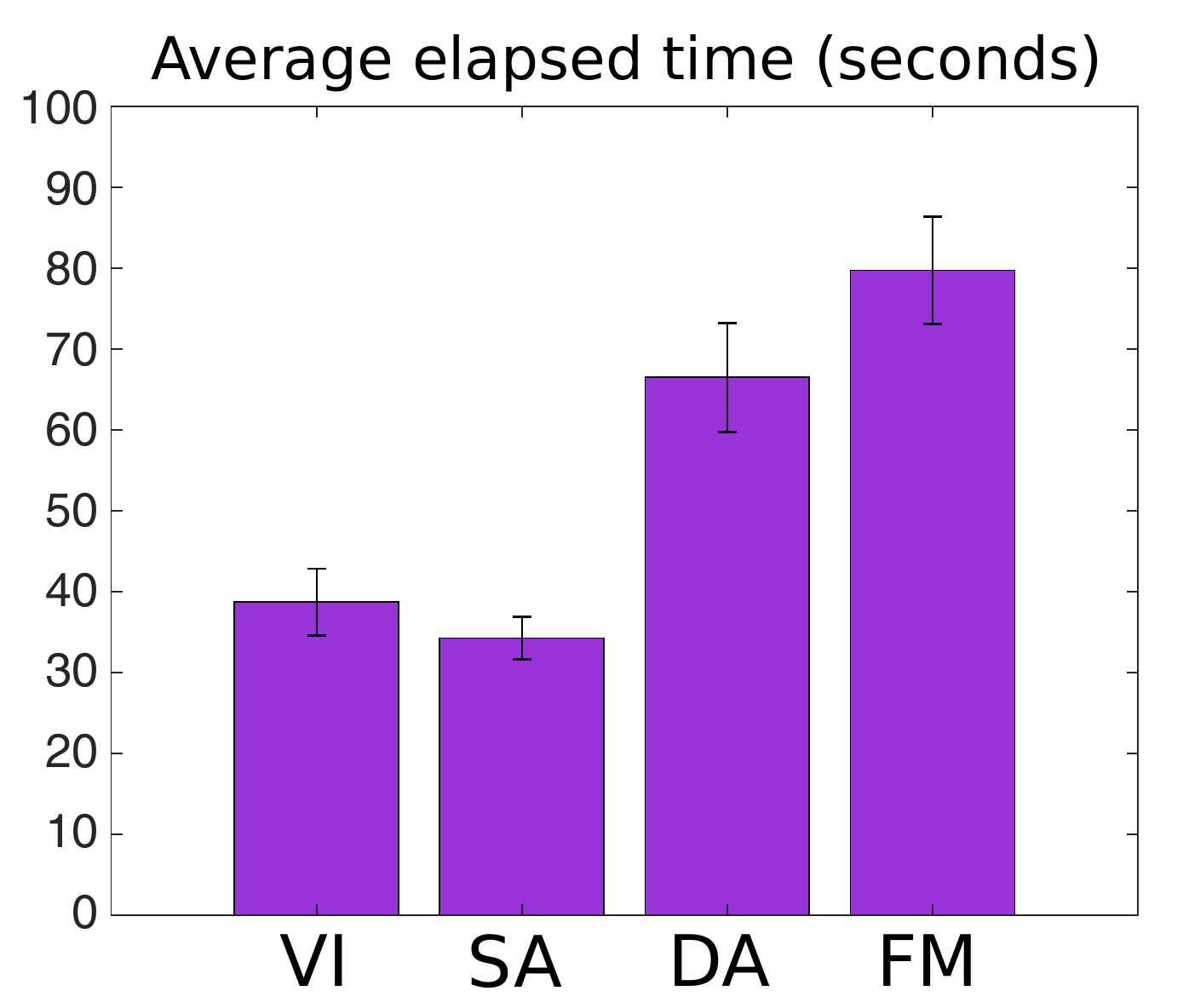}
	\caption{Average elapsed time in each material stimuli per experimental condition across all subjects. There is a significant effect of the presentations in the interaction time and, hence, in the level of immersion of the material experience.}
	\label{figure:time-mean}
\end{figure}

Given the equal ability of the studied audiovisual stimuli to communicate material qualities, we investigated whether any of the conditions results in a higher degree of immersion. Our intuition is that a higher level of immersion would translate into longer interaction times with the stimuli in the respective condition. The average elapsed time in each stimuli across all participants per experimental condition is presented in Figure \ref{figure:time-mean}. In order to test the significancy of the results, we performed repeated measures analysis of variance (ANOVA), confirming the effect of the four considered presentations on the interaction time $ [F(3,54)=15.16, \, p<0.001] $. Post-hoc tests using Bonferroni corrections revealed that the time spent by the participants was significantly higher for the DA and FM conditions. The similarity of the users' ratings for the three visual and audiovisual conditions (VI, SA and DA) reported in the previous analysis together with the fact that interacting with the real materials (FM), representing the highest possible level of immersion, conveyed longer interaction times, dismisses the possibility that this effect may be due to the complexity of the experimental task.

%% file: table-pca.tex
\begin{table*}
\centering
\scriptsize
\begin{tabular}{lrrr|rrr|rrr|rrr}
\toprule
& \multicolumn{3}{c|}{Visual} & \multicolumn{3}{c|}{Static audiovisual} & \multicolumn{3}{c|}{Dynamic audiovisual} & \multicolumn{3}{c}{Full-modal} \\
	 & \multicolumn{1}{c}{PC1} & \multicolumn{1}{c}{PC2} & \multicolumn{1}{c|}{PC3} & \multicolumn{1}{c}{PC1} & \multicolumn{1}{c}{PC2} & \multicolumn{1}{c|}{PC3} & \multicolumn{1}{c}{PC1} & \multicolumn{1}{c}{PC2} & \multicolumn{1}{c|}{PC3} & \multicolumn{1}{c}{PC1} & \multicolumn{1}{c}{PC2} & \multicolumn{1}{c}{PC3} \\
\midrule
rough--smooth & $-0.235$ & \color{myred} $\mathbf{-0.351}$ & $0.177$ & $-0.008$ & \color{myred} $\mathbf{0.483}$ & $0.161$ & $0.001$ & \color{myred} $\mathbf{0.445}$ & $0.191$ & $-0.163$ & \color{myred} $\mathbf{0.403}$ & $0.132$ \\
hard--soft & $-0.233$ & $-0.063$ & $-0.106$ & $0.115$ & $0.258$ & $-0.173$ & $0.001$ & $0.254$ & \color{myred} $\mathbf{0.535}$ & $-0.125$ & $-0.165$ & $0.306$ \\
thick--thin & \color{myred} $\mathbf{-0.390}$ & $0.065$ & \color{myred} $\mathbf{0.359}$ & $-0.265$ & $0.182$ & $0.344$ & $-0.329$ & $0.241$ & $-0.033$ & \color{myred} $\mathbf{-0.466}$ & $0.179$ & \color{myred} $\mathbf{0.372}$ \\
stiff--flexible & $-0.301$ & $-0.112$ & $0.052$ & $-0.036$ & $0.255$ & $-0.028$ & $-0.116$ & $0.297$ & \color{myred} $\mathbf{0.383}$ & $-0.192$ & $-0.085$ & $0.154$ \\
shiny--matte & \color{myred} $\mathbf{0.404}$ & $-0.090$ & $0.065$ & \color{myred} $\mathbf{0.395}$ & $-0.188$ & $-0.037$ & \color{myred} $\mathbf{0.417}$ & $-0.052$ & $-0.157$ & \color{myred} $\mathbf{0.418}$ & \color{myred} $\mathbf{-0.387}$ & \color{myred} $\mathbf{0.372}$ \\
bright--dark & \color{myred} $\mathbf{0.478}$ & $-0.134$ & \color{myred} $\mathbf{0.753}$ & \color{myred} $\mathbf{0.599}$ & $-0.178$ & \color{myred} $\mathbf{0.657}$ & \color{myred} $\mathbf{0.584}$ & $0.139$ & \color{myred} $\mathbf{0.351}$ & $0.213$ & $-0.073$ & \color{myred} $\mathbf{0.664}$ \\
transparent--opaque & \color{myred} $\mathbf{0.361}$ & \color{myred} $\mathbf{0.364}$ & $-0.265$ & \color{myred} $\mathbf{0.521}$ & $0.038$ & $-0.330$ & \color{myred} $\mathbf{0.522}$ & $0.034$ & $-0.005$ & \color{myred} $\mathbf{0.508}$ & $0.131$ & $-0.242$ \\
homog.--heterog. & $0.262$ & $0.041$ & \color{myred} $\mathbf{-0.397}$ & $0.131$ & $-0.080$ & \color{myred} $\mathbf{-0.466}$ & $0.209$ & $-0.184$ & $0.059$ & $0.106$ & \color{myred} $\mathbf{-0.390}$ & $-0.006$ \\
expensive--cheap & $0.011$ & \color{myred} $\mathbf{0.446}$ & $-0.044$ & $-0.233$ & \color{myred} $\mathbf{-0.390}$ & $-0.043$ & $-0.105$ & \color{myred} $\mathbf{-0.360}$ & $0.337$ & $-0.274$ & \color{myred} $\mathbf{-0.381}$ & $-0.027$ \\
natural--synthetic & $-0.033$ & \color{myred} $\mathbf{0.451}$ & $0.151$ & $-0.220$ & $-0.279$ & $0.272$ & $0.178$ & $-0.283$ & \color{myred} $\mathbf{0.431}$ & \color{myred} $\mathbf{-0.356}$ & $-0.310$ & $-0.159$ \\
beautiful--ugly & $0.141$ & \color{myred} $\mathbf{0.403}$ & $-0.036$ & $-0.070$ & \color{myred} $\mathbf{-0.351}$ & $-0.072$ & $-0.020$ & \color{myred} $\mathbf{-0.437}$ & $0.277$ & $-0.088$ & \color{myred} $\mathbf{-0.349}$ & $-0.120$ \\
pleasant-unpleasant & $-0.189$ & \color{myred} $\mathbf{-0.361}$ & $0.024$ & $0.063$ & \color{myred} $\mathbf{0.418}$ & $0.050$ & $-0.061$ & \color{myred} $\mathbf{0.364}$ & $-0.022$ & $0.065$ & $0.290$ & $0.221$ \\
\midrule
Explained variance [\%] & $41.02$ & $28.75$ & $12.70$ & $41.35$ & $25.94$ & $12.46$ & $38.35$ & $28.64$ & $13.13$ & $60.97$ & $17.53$ & $8.69$ \\
Cumulative variance [\%] & $41.02$ & $69.78$ & $82.49$ & $41.35$ & $67.29$ & $79.76$ & $38.35$ & $67.01$ & $80.15$ & $60.97$ & $78.51$ & $87.20$ \\  
\bottomrule
\end{tabular}
\caption{Factor loadings, explained variance and cumulative variance of the first three principal components for each condition. Bold, red values represent the strongest factors (greater than $ 0.35 $) for each principal component.}  
\label{table:factor-loadings}
\end{table*}

%% file: discussion.tex
\section{Discussion and future work}


After a comprehensive analysis of the collected user data, the primary finding of the present study is that the auditory cues employed in our studies do not contribute with additional value to the perception of material qualities. All of the conducted evaluations indicate that the three digital conditions evaluated (VI, SA and DA) have a fairly equal ability transmit material information without weakening the overall experience. The most plausible explanation for this outcome may be that the utilization of contact sounds from rubbing interactions exclusively did not yield enough information to discriminate between the two different material classes or to characterize specimens within the same class. Indeed, previous investigations concerning the perception of textiles have asessed other gestures like two-finger pinching, stroking or sample scrunching as the most repeated interactions when evaluating real fabric samples \cite{Atkinson2013}. Another reason that may have influenced our results is the fact that the employed material classes (leathers and fabrics) do not differ significantly when considering their characteristic sounds. Although our research aimed at examining the perceived intra-class differences between materials, it has been documented that not even striking sounds provide sufficient cues to differentiate samples within the same class \cite{Giordano2006}.

Another interesting finding concerns the ability to discriminate between leathers and fabrics through the experimental conditions. Both the PCA analysis and the SVM classification indicate that it is possible to discern between these two classes based on the ratings for the selected set of attributes, when the real materials are provided. However, this capacity is not translated well to the conditions where only images and sound from the material samples are provided. As regards to which types of qualities allow better material discrimination, considering the visual quality features alone provide the highest accuracies. Interestingly, the presence affective features has certain influence in the digital conditions (VI, SA and DA) in comparison to tactile qualities, which are more salient when the real materials are provided. This supports our intuition that affective properties have a meaningful role in the perception of digital products.

During informal interviews after the realization of the experimental task, subjects reported to have enjoyed the utilization of the tactile sonification interface. This reported engagement translated into significantly longer interaction times with the system during the DA condition, $ 66.5 $ seconds of average interaction per material, in contrast to the $ 38.7 $ and $ 34.3 $ seconds on average for the VI and SA conditions respectively. This is in accordance with the experimental results from Ho et al. \cite{Ho2013}, where the addition of realistic auditory feedback led to considerably longer ($ 30\% $) interplays with their AR system. Moreover, the similarity of the ratings between all these three conditions dismisses the possibility that such effect is due to the complexity of the task. These results, however, must be viewed with some caution as more exhaustive experiments should be conducted in order to further explore this insight. Indeed, the investigation of how the presence of interactive sounds affects the level of immersion and engagement when exploring digital materials remains a promising avenue for future research.

Future investigations may turn over to other kinds of touch-related material sounds, more consistent with the actual human behavior, for which alternative synthesis approaches could be more suited. For instance, physically-based synthesis methods have been able to generate the distinctive crumpling sound of materials \cite{Cirio2016} at, however, unfeasible computation times. Deep learning techniques could also leverage the synthesis of contact sounds \cite{Owens2016}, provided that a sufficiently rich database of sounds is given for the training of the model.

\section{Conclusion}

The main purpose of this investigation is to determine the impact of an interactive material sonification system in the perception of physical and affective material qualities. For the development of this sonification algorithm, we relied on granular synthesis to interactively reproduce characteristic contact sounds generated when rubbing leather and fabric materials with the fingertip. This method, which has been specifically developed for tactile devices, plays back chunks of sound (grains) upon tactile interaction with the material images on the screen. Its performance has been then evaluated by examining its ability to describe concrete material qualities in contrast to additional visual, audiovisual and full-modal conditions, via a psychophysical study. We discovered that the contact sounds employed in our experiment do not contribute with additional information to the perception of material qualities, since all the considered digital condition exhibit almost similar performance. Furthermore, we observed that our sonification method has a significant effect in the users' immersive experience when interacting with digital materials. In light of these findings, we provide several potential lines of future research regarding sonification systems for digital materials, which may include additional material categories, more suitable types of interaction and alternative synthesis techniques.

%% file: main-arxiv.bbl

\begin{thebibliography}{00}


\ifx \showCODEN    \undefined \def \showCODEN     #1{\unskip}     \fi
\ifx \showDOI      \undefined \def \showDOI       #1{{\tt DOI:}\penalty0{#1}\ }
  \fi
\ifx \showISBNx    \undefined \def \showISBNx     #1{\unskip}     \fi
\ifx \showISBNxiii \undefined \def \showISBNxiii  #1{\unskip}     \fi
\ifx \showISSN     \undefined \def \showISSN      #1{\unskip}     \fi
\ifx \showLCCN     \undefined \def \showLCCN      #1{\unskip}     \fi
\ifx \shownote     \undefined \def \shownote      #1{#1}          \fi
\ifx \showarticletitle \undefined \def \showarticletitle #1{#1}   \fi
\ifx \showURL      \undefined \def \showURL       #1{#1}          \fi

\bibitem{An2012}
{S. An}, {D. James}, {and} {S. Marschner}. 2012.
\newblock \showarticletitle{Motion-driven Concatenative Synthesis of Cloth
  Sounds}.
\newblock {\em ACM Trans. on Graphics\/} {31}, 4 (July 2012), 102:1--102:10.
\newblock
\showISSN{0730-0301}


\bibitem{Atkinson2013}
{D. Atkinson}, {P. Orzechowski}, {B. Petreca}, {N. Bianchi-Berthouze}, {P.
  Watkins}, {S. Baurley}, {S. Padilla}, {and} {M. Chantler}. 2013.
\newblock \showarticletitle{Tactile Perceptions of Digital Textiles: A Design
  Research Approach}. In {\em Proc. of the SIGCHI Conf. on Human Factors in
  Computing Systems} {\em (CHI '13)}. ACM, 1669--1678.
\newblock
\showISBNx{978-1-4503-1899-0}


\bibitem{Barrass2002}
{S. Barrass} {and} {M. Adcock}. 2002.
\newblock \showarticletitle{Interactive Granular Synthesis of Haptic Contact
  Sounds}. In {\em 22nd Audio Engineering Society Conf.: Virtual, Synthetic,
  and Entertainment Audio}. Audio Engineering Society.
\newblock


\bibitem{Brinkmann2012}
{P. Brinkmann}. 2012.
\newblock {\em Making Musical Apps}.
\newblock O'Reilly Media, Inc.
\newblock


\bibitem{Cirio2016}
{G. Cirio}, {D. Li}, {E. Grinspun}, {M.A. Otaduy}, {and} {C. Zheng}. 2016.
\newblock \showarticletitle{Crumpling Sound Synthesis}.
\newblock {\em ACM Trans. on Graphics\/} {35}, 6, Article 181 (Nov. 2016), 11
  pages.
\newblock
\showISSN{0730-0301}


\bibitem{Citrin2003}
{A. Citrin}, {D.~Stem Jr.}, {E. Spangenberg}, {and} {M. Clark}. 2003.
\newblock \showarticletitle{Consumer need for tactile input: An internet
  retailing challenge}.
\newblock {\em Journal of Business Research\/} {56}, 11 (2003), 915 -- 922.
\newblock
\showISSN{0148-2963}


\bibitem{Etzi2014}
{R. Etzi}, {C. Spence}, {and} {A. Gallace}. 2014.
\newblock \showarticletitle{Textures that we like to touch: An experimental
  study of aesthetic preferences for tactile stimuli}.
\newblock {\em Consciousness and Cognition\/}  {29} (2014), 178--188.
\newblock
\showISSN{1053-8100}


\bibitem{Fleming2013}
{R.W. Fleming}, {C. Wiebel}, {and} {K. Gegenfurtner}. 2013.
\newblock \showarticletitle{Perceptual qualities and material classes}.
\newblock {\em Journal of Vision\/} {13}, 8 (2013), 9.
\newblock


\bibitem{Fuchs1980}
{H. Fuchs}, {Z.M. Kedem}, {and} {B.F. Naylor}. 1980.
\newblock \showarticletitle{On Visible Surface Generation by a Priori Tree
  Structures}.
\newblock {\em SIGGRAPH Comput. Graph.\/} {14}, 3 (July 1980), 124--133.
\newblock
\showISSN{0097-8930}


\bibitem{Fujisaki2015}
{W. Fujisaki}, {M. Tokita}, {and} {K. Kariya}. 2015.
\newblock \showarticletitle{Perception of the material properties of wood based
  on vision, audition, and touch}.
\newblock {\em Vision Research\/}  {109} (2015), 185 -- 200.
\newblock
\showISSN{0042-6989}


\bibitem{GarridoJurado2014}
{S. Garrido-Jurado}, {R. Muñoz-Salinas}, {F.J. Madrid-Cuevas}, {and} {M.J.
  Mar\'in-Jim\'enez}. 2014.
\newblock \showarticletitle{Automatic generation and detection of highly
  reliable fiducial markers under occlusion}.
\newblock {\em Pattern Recognition\/} {47}, 6 (2014), 2280 -- 2292.
\newblock
\showISSN{0031-3203}


\bibitem{Giordano2006}
{B.L. Giordano} {and} {S. McAdams}. 2006.
\newblock \showarticletitle{Material identification of real impact sounds:
  Effects of size variation in steel, glass, wood, and plexiglass plates}.
\newblock {\em The Journal of the Acoustical Society of America\/} {119}, 2
  (2006), 1171--1181.
\newblock


\bibitem{Gower1975}
{J.C. Gower}. 1975.
\newblock \showarticletitle{Generalized procrustes analysis}.
\newblock {\em Psychometrika\/} {40}, 1 (01 Mar 1975), 33--51.
\newblock
\showISSN{1860-0980}


\bibitem{Guest2002}
{S. Guest}, {C. Catmur}, {D. Lloyd}, {and} {C. Spence}. 2002.
\newblock \showarticletitle{Audiotactile interactions in roughness perception}.
\newblock {\em Experimental Brain Research\/} {146}, 2 (2002), 161--171.
\newblock
\showISSN{0014-4819}


\bibitem{Ho2013}
{C. Ho}, {R. Jones}, {S. King}, {L. Murray}, {and} {C. Spence}. 2013.
\newblock \showarticletitle{Multisensory augmented reality in the context of a
  retail clothing application}. In {\em Audio Branding Academy yearbook
  2012/2013}. Nomos Publishers, Germany, 167--174.
\newblock


\bibitem{Klatzky2010}
{R. Klatzky} {and} {S. Lederman}. 2010.
\newblock \showarticletitle{Multisensory Texture Perception}.
\newblock In {\em Multisensory Object Perception in the Primate Brain}.
  Springer New York, 211--230.
\newblock
\showISBNx{978-1-4419-5615-6}


\bibitem{Klatzky2000}
{R.L. Klatzky}, {D.K. Pai}, {and} {E.P. Krotkov}. 2000.
\newblock \showarticletitle{Perception of material from contact sounds}.
\newblock {\em Presence: Teleoperators and Virtual Environments\/} {9}, 4
  (2000), 399--410.
\newblock


\bibitem{Martin2015}
{R. Mart{\'i}n}, {J Iseringhausen}, {M. Weinmann}, {and} {M.B. Hullin}. 2015.
\newblock \showarticletitle{Multimodal Perception of Material Properties}. In
  {\em ACM Symposium on Applied Perception} {\em (SAP '15)}. ACM, 33--40.
\newblock
\showISBNx{978-1-4503-3812-7}


\bibitem{Martin2017}
{R. Mart{\'i}n}, {M. Weinmann}, {and} {M.B. Hullin}. 2017.
\newblock \showarticletitle{Digital Transmission of Subjective Material
  Appearance}.
\newblock {\em Journal of WSCG\/} {25}, 2 (June 2017), 57--66.
\newblock
\showISSN{1213-6972}


\bibitem{Misra2009}
{A. Misra} {and} {P. Cook}. 2009.
\newblock \showarticletitle{Toward Synthesized Environments: A Survey of
  Analisis and Synthesis Methods for Sound Designers and Composers}. In {\em
  Proc. of the International Computer Music Conf. (ICMC)}. Int. Computer Music
  Association, 155--162.
\newblock


\bibitem{Owens2016}
{A. Owens}, {P. Isola}, {J. McDermott}, {A. Torralba}, {E.~H. Adelson}, {and}
  {W.~T. Freeman}. 2016.
\newblock \showarticletitle{Visually Indicated Sounds}. In {\em IEEE Conf. on
  Computer Vision and Pattern Recognition (CVPR)}. 2405--2413.
\newblock


\bibitem{Spence2006}
{C. Spence} {and} {M. Zampini}. 2006.
\newblock \showarticletitle{Auditory contributions to multisensory product
  perception}.
\newblock {\em Acta Acustica united with Acustica\/} {92}, 6 (2006),
  1009--1025.
\newblock


\bibitem{Tajadura-Jimenez2014}
{A. Tajadura-Jim{\'e}nez}, {B. Liu}, {N. Bianchi-Berthouze}, {and} {F.
  Bevilacqua}. 2014.
\newblock \showarticletitle{Using Sound in Multi-touch Interfaces to Change
  Materiality and Touch Behavior}. In {\em Proc. of the 8th Nordic Conf. on
  Human-Computer Interaction} {\em (NordiCHI '14)}. ACM, 199--202.
\newblock
\showISBNx{978-1-4503-2542-4}


\end{thebibliography}
